\documentclass[12pt, prd, showpacs]{revtex4}
\usepackage{amsmath}

\begin{document}

\title{Entropy of quasiblack holes}
\author{Jos\'{e} P. S. Lemos}
\affiliation{Centro Multidisciplinar de Astrof\'{\i}sica - CENTRA,
Departamento de F\'{\i}sica, 
Instituto Superior T\'ecnico - IST, Universidade T\'{e}cnica de Lisboa
- UTL, Avenida Rovisco Pais 1, 1049-001 Lisboa, Portugal\,\,}
\email{joselemos@ist.utl.pt}
\author{Oleg B. Zaslavskii}
\affiliation{Astronomical Institute of Kharkov V.N. Karazin National
University, 35
Sumskaya Street, Kharkov, 61022, Ukraine\,\,}
\email{ozaslav@kharkov.ua}

\begin{abstract}
We trace the origin of the black hole entropy $S$ replacing a black hole
by a quasiblack hole. Let the boundary of a static body approach its own
gravitational radius, in such a way that a quasihorizon forms. We show
that if the body is thermal with the temperature taking the Hawking
value at the quasihorizon limit, it follows, in the nonextremal case,
from the first law of thermodynamics that the entropy approaches the
Bekenstein-Hawking value $ S=A/4$. In this setup, the key role is played
by the surface stresses on the quasihorizon and one finds that the
entropy comes from the quasihorizon surface. Any distribution of matter
inside the surface leads to the same universal value for the entropy in
the quasihorizon limit. This can be of some help in the understanding of
black hole entropy. Other similarities between black holes and
quasiblack holes, such as the mass formulas for both objects had been
found previously. We also discuss the entropy for extremal quasiblack
holes, a more subtle issue.
\end{abstract}

\maketitle





\section{Introduction}

As it is known, the entropy $S$ of a nonextremal black hole is equal to the
Bekenstein-Hawking value, $S=A/4$, where $A$ is the area of the black hole
horizon (we use units such that Newton's constant, Planck's constant, and
the speed of light are put to one). Its formal appearance is especially
transparent in a Euclidean action approach where this term stems entirely
due to the presence of the horizon \cite{gh,y1,y2,oz}. A development on
these issues was performed by Brown and York \cite{by}, where from a
quasilocal energy formalism one can deduce, among other things, black
hole thermodynamics itself.

Imagine now, a collapsing body. When the surface of the body is close to its
own horizon $r_{+}$, but does not coincide with it, there is no obvious
reason for the presence of such an $S=A/4$ term for the entropy of the body.
Therefore, at first glance, the entropy $A/4$ appears as a jump, when the
black hole forms. Nonetheless, we will see below that we can restore the
continuity and trace the origin of the entropy under discussion, if instead
of a black hole we will consider a quasiblack hole. Roughly speaking, a
quasiblack hole is an object in which the boundary gets as close as one
likes to the horizon. However, an event horizon does not form (see \cite
{qbh,static,statio} and references therein for more on the definition and
properties of quasiblack holes, and \cite{lw1,lw2,exdust,klz,jmp,mod,nod}
for examples of quasiblack holes themselves).

We will study a distribution of matter constrained inside a boundary, at
some temperature $T$, in the vicinity of being a quasiblack hole and
find the entropy $S$ of such a system. The steps to find the universal
entropy formula consist of using the first law together with the
quasilocal formalism of \cite{by}. We will see that the matter entropy
helps in the generation of the $A/4$ term when the boundary approaches
the horizon. Instead of giving some examples of calculating the entropy
for some spherically symmetric shell (see, e.g, the very interesting
studies in \cite {page,mar}), we proceed in a model-independent way and
exploit essentially the fact that the boundary almost coincides with the
would-be horizon, i.e, with the quasihorizon. In the procedure, it is
essential that some components of surface stresses diverge in the
quasihorizon limit, in the case of nonextremal quasiblack holes. It is
the price paid for keeping a shell (which is inevitable near the
nonextremal quasihorizon \cite{qbh}) in equilibrium without collapsing.
Although by themselves, such stresses that grow unbounded look
unphysical, the whole quasiblack hole picture turns out to be at least
useful methodically since it enables us to trace some features of black
holes. In particular, in previous works \cite {static,statio} we managed
to derive the black hole mass formula using a quasiblack hole approach
in which both the meaning of terms and their derivation are different
from the standard black hole case. Thus, the concept of a quasiblack
hole has two sides: it simultaneously mimics some features of black
holes but also shows those features in a quite different setting. To
some extent, it can be compared with the membrane paradigm
\cite{membr,pw}. However, the key role that the diverging stresses play
here, was not exploited there (a more detailed general comparison of
both approaches would be important, but is beyond the scope of our
paper). In the present work, we extend the approach to the entropy of
nonextremal quasiblack holes. We also discuss the issue of the entropy
for extremal quasiblack holes. In some papers \cite{haw,teit,gk} it was
argued that $S=0$ for extremal black holes, while in other works there
were arguments to recover the value $S=A/4$ (see \cite{and} for works
within general relativity, and \cite{fn} for additional arguments within
string theory). Thus, it is useful to examine the issue of the entropy
of extremal quasiblack holes at the classical level.

A point worth raising is that the entropy of a pure black hole can in part
be recovered from entanglement arguments \cite{bom,sred}. Quasiblack holes
appeared first in the context of self-gravitating magnetic monopoles \cite
{lw1,lw2} and this prompted Lue and Weinberg \cite{lueweinberg3} to argue
that an observer in the outer region describes the quasiblack hole inner
region in terms of a statistical density matrix $\rho $ defined upon taking
the trace of the degrees of freedom in this inner region. Then, defining as
usual the entropy $S$ as $S=-\mathrm{Tr}\,\left( \rho \,\ln \rho \right) $,
one can argue that the emergence of the black hole entropy, or quasiblack
hole entropy, can be consistently ascribed to the entanglement of the outer
and inner fields \cite{lueweinberg3}. This entanglement entropy has some
drawbacks, one of which is that, although it gives a quantity proportional
to the horizon area $A$, the proportionality constant is infinite, unless
there is a ultraviolet cutoff presumably at the Planck scale, which somehow
would give in addition the required 1/4 value. Here we do not touch on the
physical statistical approach, we rather use the thermodynamic approach to
the entropy of quasiblack holes and find precisely the value $S=A/4$.

The issue discussed by us is also relevant in the context of \cite{jac} (see
also \cite{br,par,pad}), where it was argued that the Einstein equation can
be derived from the first law of thermodynamics $T\,dS=\delta Q$ and the
proportionality between the entropy and the area $A$ of some causal horizon.
In this setting, $T$ and $\delta Q$ are the temperature and the heat flux
seen by an accelerated local Rindler observer on a surface which is close to
the horizon but does not coincide with it and remains timelike. Since this
setup is based on a timelike surface, rather than on a null surface,
strictly speaking there is a gap in the derivation, as the $S=A/4$ value
follows from the space-time structure for the horizon, a null surface. In
the present work we show how this $S=A/4$ term is indeed recovered in the
quasihorizon limit and, thus, fill this gap.

\section{Entropy and first law of thermodynamics for quasiblack holes}

\subsection{Entropy in the nonextremal case}

\subsubsection{Entropy formula}

Consider a static metric, not necessarily spherically symmetric. We assume
that there is a compact body. Then, at least in some vicinity of its
boundary the line element can be written in Gaussian coordinates as
\begin{equation}
ds^{2}=-N^{2}dt^{2}+dl^{2}+g_{ab}\,dx^{a}dx^{b}\,,  \label{m}
\end{equation}
where $(t,l)$ are the time and radial coordinates, respectively, and $
x^{a},x^{b}$ represent the angular part. We suppose that the boundary of the
compact body is at $l=\mathrm{\ const}$. The metric functions $N$ and $
g_{ab} $ generically have different forms for the inner and outer parts. Now
also assume that the system is at a local Tolman temperature $T$ given by
\begin{equation}
T=\frac{T_{0}}{N}\,,  \label{tol}
\end{equation}
where $T_{0}=\mathrm{const}$. $T_{0}$ should be considered as the
temperature at asymptotically flat infinity. Assuming the validity of the
first law of thermodynamics we can write it in terms of boundary values
\begin{equation}
Td(s\sqrt{g})=d(\sqrt{g}\epsilon )+\frac{\Theta ^{ab}}{2}\sqrt{g}
\,dg_{ab}+\varphi\, d(\sqrt{g}\,\rho_{\mathrm{e}})\,\,.  \label{fl}
\end{equation}
One should carefully define each term appearing in Eq.~(\ref{fl}). The
quantity $g$ is defined as $g\equiv \det g_{ab}$. The quantity $s$ is the
entropy density entering the expression for the total entropy
\begin{equation}
S=\int d^{2}x\sqrt{g}\,s\,.  \label{e1}
\end{equation}
The quantity $\epsilon $ is the quasilocal energy density, defined as \cite
{by}
\begin{equation}
\epsilon =\epsilon ^{\mathrm{g}}-\epsilon ^{0}  \label{k}
\end{equation}
where
\begin{equation}
\epsilon ^{\mathrm{g}}=\frac{K}{8\pi }  \label{gk1}
\end{equation}
and $\epsilon ^{0}=K^{0}/8\pi $,. So, $\epsilon =\frac{K-K_{0}}{8\pi }$.
Here $K$ is the trace of the two-dimensional extrinsic curvature $K_{ab}$ of
the boundary surface embedded in the three-dimensional manifold $t=\mathrm{\
const}$, and $K_{0}$ is a similar term for the reference background
manifold, e.g., flat spacetime, but in our context the precise background is
unimportant. In more detail,
\begin{equation}
K_{ab}=-\frac{1}{2}\,{g_{ab}}^{\,\prime }\,,
\end{equation}
is the extrinsic curvature, where a prime denotes differentiation with
respect to $l$, i.e., ${}^{\prime }\equiv \frac{\partial \,}{\partial l}$,
and
\begin{equation}
K=-\frac{1}{\sqrt{g}}\,\sqrt{g}^{\,\,\prime }
\end{equation}
its trace, $K=K_{ab}\,g^{ab}$. Finally, the spatial energy-momentum tensor
$
\Theta _{ab}$ is equal to \cite{by} (see also \cite{isr,mtw,boothm} for the
more traditional approach),
\begin{equation}
\Theta _{ab}=\Theta _{ab}^{g}-\Theta _{ab}^{0}  \label{sab}
\end{equation}
where
\begin{equation}
8\pi \Theta _{ab}^{g}=K_{ab}+\left( \frac{N^{\prime }}{N}-K\right) g_{ab}\,,
\label{p}
\end{equation}
and $\Theta _{ab}^{0}$ is the corresponding background tensor, with a form
similar to Eq.~(\ref{p}). Finally, the quantity $\varphi $ is the electric
potential and $\rho _{\mathrm{e}}$ is the electric charge density of the
matter.

Our strategy consists of integrating the first law (\ref{fl}) to
obtain the entropy. In general, for an arbitrary boundary, if one is
far from the quasihorizon, the integration procedure requires
knowledge of the equation of state of the matter, as has been worked
out by Martinez for some specific models of a shell in vacuum
\cite{mar}. However, we will now see that if we choose a sequence of
configurations such that all its members remain on the threshold of
the formation of a horizon, and integrate just over this very subset,
the answer turns out to be model independent, and there is no need to
specify an equation of state. To this end, we must 
simultaneously change 
the size and the proper mass $M$ of the configuration,
to keep it near its gravitational radius $r_{+}$, and in such a way
that $N\rightarrow 0 $ for all such configurations. This will allow us
to integrate the first law along such a sequence and obtain the value
of the entropy for a shell near the quasihorizon.

We point out 
some subtleties. We use the definitions of quasiblack holes done in
\cite{qbh,static,statio} for general and spherical systems. Consider, for
simplicity, spherical configurations. Then, there are two relevant
quantities, the system radius $R$ and its gravitational radius $r_{+}$. For
$
r_{+}$ fixed and $R\rightarrow r_{+}$, we deal with the situation of \cite
{qbh,static,statio}. In doing $N\rightarrow 0$ the small parameter is $
\varepsilon =\frac{R-r_{+}}{R}\ll 1$. But, now, actually there are two
small parameters $\varepsilon $ and $\delta=\frac{(\delta
r_{+})}{r_{+}}$ since we want to consider small variations of
thermodynamic quantities for two close systems which differ in
$r_{+}$. Then, we must first send $\varepsilon \rightarrow 0$ and only
afterward consider $\delta \ll 1$. It is this approach that ensures
that we are dealing with quasiblack holes having slightly different
radii $r_{+}$. If the system is kept near the quasihorizon, this
allows us to integrate the first law along such a sequence counting
different members of the same family of states and obtain the value of
$S$ for the shell layer at the quasihorizon. It is worth dwelling on
this point, and in order to understand it, consider, for instance, the
simplest configuration with a Schwarzschild exterior solution and a
Minkowski metric inside. Then, the Arnowitt-Deser-Misner (ADM) mass
$m$ of the solution, the radius of the shell $R$ and its proper mass
$M$ are connected by the relation $m=M- \frac{M^{2}}{2R}$
\cite{y1,page}. The horizon, if there is one, is at radius
$r_{+}\equiv2m$. We can characterize the system by two independent
parameters, $R$ and $m$, say. Then, in the whole space of parameters,
we must choose the curve lying slightly above the straight line
$R=2m$. Then, in the process of integration along this curve all three
quantities $R$, $m$ and $M$ change but in such a way that the
approximate equality $R\approx 2m$ holds.

Thus, in the outer region, neglecting the difference between quasiblack hole
and black hole metrics (which can be done), we can write \cite{vis0,vis} for
the system near the formation of a quasihorizon,
\begin{equation}
K_{ab}=K_{ab}^{(1)}l+O(l^{2})\,.  \label{reg}
\end{equation}
Equation~(\ref{reg}) follows from regularity conditions. It is then seen from 
(\ref{k}) that the quasilocal energy density $\epsilon $ remains finite, and
from (\ref{p}) that the spatial stresses $\Theta _{ab}$ diverge due to the
term $\frac{1}{N}\,\left( \frac{\partial N}{\partial l}\right) _{+}$. In the
outer region \ $\left( \frac{\partial N}{\partial l}\right) _{+}\rightarrow
\kappa $ where $\kappa $ is the surface gravity of the body. Leaving in Eq.
(\ref{fl}) only the dominant contribution, we obtain that
\begin{equation}
d(s\sqrt{g})=\frac{\kappa }{16\pi T_{0}}\sqrt{g}g^{ab}dg_{ab}\,.  \label{ds}
\end{equation}
Up to now, the quantity $T_{0}$ is arbitrary. However, we should take into
account that near the quasihorizon quantum fields are inevitably present.
For an arbitrary temperature, their backreaction becomes divergent and only
the choice
\begin{equation}
T_{0}=T_{\mathrm{H}}=\frac{\kappa }{2\pi }\,,  \label{kawktemp}
\end{equation}
where $T_{\mathrm{H}}$ is the Hawking temperature, enables us to obtain a
finite result (see, e.g., \cite{and} and references therein, for the proof
that the stress-energy tensor and other quantities diverge strongly on the
horizon unless the corresponding fields are in a state with a temperature
equal to the natural black hole temperature $T_{\mathrm{H}}$; we list the
corresponding expression for the stress-energy tensor below in 
Eq.~(\ref{set})). 
If, thus, neglecting again the difference between a black hole and
quasiblack hole, we substitute this equality in (\ref{ds}), we obtain
\begin{equation}
d(s\sqrt{g})=\frac{1}{4}\,\,d\sqrt{g}\,.
\end{equation}
Upon integration over an area $A$, we reproduce the famous result
\begin{equation}
S=\frac{1}{4}\,A\,,  \label{4}
\end{equation}
up to a constant $c$. In these considerations, we took into account the
leading term while integrating the first law (\ref{fl}). It follows from the
Tolman formula (\ref{tol}) that the corrections which come from the first
and last terms are of the order $O(\sqrt{-g_{00}})$ and vanish when the
quasihorizon is approached. In general, there are also corrections which
stem from the second term. They are model dependent and depend on how
rapidly the temperature $T_{0}$ approaches the Hawking value
$T_{\mathrm{H}}$. 
In the limit $T_{0}\rightarrow T_{\mathrm{H}}$ they vanish by
construction.

It is interesting to note that in the quasihorizon limit an analog of the
Euler relation is found. Indeed, the Euler relation has the form
\begin{equation}
Ts=\sigma +p\,,  \label{eu}
\end{equation}
where $\sigma $ is the energy density and $p$ plays the role of a pressure.
It is easy to check that the relation (\ref{eu}) does not hold in general.
However, one can check directly that near the quasihorizon Eq.~(\ref{eu})
with the mean pressure given by $p={\Theta ^{ab}g_{ab}}/{2}$ holds
approximately. In doing so, in the main approximation, one finds $p\approx
{\kappa}/{8\pi N}$ and 
$s\approx \frac{1}{4}$, in agreement with (\ref{4}).
One sees that $\sigma $ does not enter this equation at all, being thus
negligible.

\subsubsection{Choice of the constant}

We obtained that in the quasiblack hole limit $S=A/4+c$, where $
c=\mathrm{const.}$ To substantiate our choice $c=0$, we can require
that $S\rightarrow 0$ when the quasiblack hole disappears, so $A\rightarrow
0 $. Then, indeed $c=0$ and the continuity of the quasiblack hole
entropy ensues.

\subsubsection{Layer and boundary stresses}

\noindent (i) Universality of the entropy formula, independently of the
specific layer or boundary stresses adopted for the model

\noindent Our derivation is essentially based on the quasilocal approach
\cite{by} which also admits an interpretation in terms of the formalism of
thin shells \cite{isr,mtw} (see \cite{boothm} for a discussion of this point
and a more general setup, where mass and quasilocal energy are defined for a
naked black hole, a relative of a quasiblack hole). In our context this is
especially important although it does not show up quite explicitly.

Equation (\ref{sab}) for the surface stresses refers to the difference
between quantities defined in the given metric and those coming from
the background metric, say, a flat one. So Eq.~(\ref{sab}) does not
necessarily require a thin shell as a model for the matter, it is
irrelevant for the calculation of quasilocal quantities in the outer
region, so this equation is valid for any inner region joined through
the boundary to an outer region (say, to vacuum) with fixed boundary
data. This is because information about the inner region is encoded in
the boundary values \cite{y1,y2,oz}. For example, denoting by $r$ the
radial coordinate, if there is a spherically symmetric body with
radius $R$, such that $0\leq r\leq R$, and with a distribution of
matter with a mass function $m(r)$, the quasilocal energy at some $r$,
$E(r)$, is $E(r)=r\left( 1-\sqrt{1-\frac{2m({r})}{{r}}} \right) $.  It
only depends on quantities defined at $r$. In particular, at the
surface, $ r=R$, $E(R)=R\left( 1-\sqrt{1-\frac{2m({R})}{R}}\right)$.

However, in any model, be it thin shell or distributed matter joined
by a boundary to the outer space, a crust in the form of a thin
layer arises inevitably when some surface behaving as the boundary of
the matter approaches the quasihorizon. Moreover, the amplitude of
such stresses becomes infinite in this limit for nonextremal
quasiblack holes \cite{qbh}.

Thus, whatever distribution of entropy a body would have, in the
quasihorizon limit all the material distributions (including a
disperse distribution or even other shells of matter) give the same
universal result $S\rightarrow \frac{A}{4}$. So the total entropy
agrees then with the Bekenstein-Hawking value. Here, in the quasiblack
hole approach we see a manifestation of the universality inherent to
black hole physics in general and black hole entropy in
particular. One reservation is in order. If we try to take into
account the thermal radiation from the boundary toward the inside
region, we encounter the difficulty that the local temperature $
T=T_{0}/N$ grows unbound, and so the mass of the radiation and the
entropy also explode. As a result, collapse ensues with the appearance
of a true black hole inside the shell instead of a quasiblack
hole. However, because of the infinite redshift due to the factor $N$,
any typical time $t_{0}$ connected with emission of photons inside
will grow unbounded as $\frac{t_{0}}{ N}$ for an external observer as
well, so the concept of a quasiblack hole remains valid and
self-consistent, up to an almost infinite time in the limit under
discussion.

\vskip 0.4cm 
\noindent (ii) nonessentiality of the choice of the background
stresses

\noindent In the above derivation of the entropy it is essential that the
boundary approaches a quasihorizon. If we have some distribution of matter,
two shells, say, in general different cases of forming a quasihorizon are
possible \cite{exdust,klz,jmp}: in one case it can appear at the outer
shell, in the other the horizon forms at the inner one. In the latter case
the derivation of the entropy formula follows the same lines as before but
with the change that the role of boundary is now played by the inner shell,
or, more specifically, by a surface on the inner shell. Thus, in any case
the term $\Theta _{ab}^{g}$ is to be calculated near the quasihorizon from
the outside.

As far as the subtraction term $\Theta _{ab}^{0}$ is concerned, one can
choose among several different possibilities. In the present work we have
chosen flat spacetime to find $\Theta _{ab}^{0}$, in previous works \cite
{static,statio} we have chosen it differently. However, this difference is
irrelevant in the given context. Indeed, one can also calculate the stresses
given in Eq.~(\ref{sab}) using a modified version of Eq.~(\ref{sab}) itself,
in which $\Theta _{ab}^{0}$ is replaced by the term $\Theta _{ab}^{-}$
determined from the inside. In general both quantities $\Theta
_{ab}^{g}-\Theta _{ab}^{0}$ and $\Theta _{ab}^{g}-\Theta _{ab}^{-}$ are
different and even refer to different boundaries. To clarify, let us suppose
the following example: one has a thick shell with an inner radius $R_{
\mathrm{in}}$ and an outer radius $R_{ \mathrm{out}}$, with $R_{\mathrm{in}
}<R_{\mathrm{out}}$. Let us also assume that when the system approaches its
own gravitational radius $r_{+}$, one has $R_{\mathrm{out}}\rightarrow R_{
\mathrm{in}}\rightarrow r_{+}$. (This is not a generic behavior, it is an
example, see \cite{exdust,klz,jmp} for other manners in which the system can
approach $r_{+}$; in such cases one should redefine the boundary). In this
example $\Theta _{ab}^{0}$ refers to the background (say, flat spacetime)
energy-momentum tensor at the outer radius $R_{\mathrm{out}}$ (which is the
radius of the outer boundary and, thus, the radius of the system as a
whole), whereas $\Theta _{ab}^{-}$ refers to the energy-momentum tensor at
the radius $R_{\mathrm{in} }$. But now note that, in our example, in the
quasihorizon limit, when $R_{ \mathrm{out}}\rightarrow R_{\mathrm{in}
}\rightarrow r_{+}$, the difference, $\Theta _{ab}^{-}-\Theta _{ab}^{0}$,
between both subtraction terms becomes inessential for our purposes. This is
because both $\Theta _{ab}^{-}$ and $\Theta _{ab}^{0}$ remain finite.
Indeed, $\Theta _{ab}^{0}$ is finite by its very meaning, and the only
potentially dangerous term in $\Theta _{ab}^{-}$, namely $\left(
N^{-1}\frac{
\partial N}{\partial l}\right) ^{-}$, is also finite since in the inner
region $N=\epsilon f(l)$ where, by definition of a quasiblack hole, $
\epsilon $ is a small parameter and $f(l)$ is a regular function (see \cite
{qbh} and especially \cite{static} for more detailed explanations). In the
calculation of the entropy described above all these finite terms are
multiplied by the factor $N\rightarrow 0$, and do not contribute. The
non-zero contribution to the entropy comes from the leading divergent term
$
\frac{1}{N}\frac{\partial N}{\partial l}\approx \frac{\kappa }{N}$ in $
\Theta _{ab}^{g}$. In addition, it is worth noting that it is this term
which ensures the existence of a mass formula for quasiblack holes similar
to the mass formula for black holes \cite{static,statio}. Now we see that,
actually, this term plays a crucial role also in the derivation of the
entropy of quasiblack holes.

The issue of the influence of boundary stresses at infinity (see,
e.g., \cite {asflat}) has been in great focus recently. Although
important, we are mainly interested in local boundary stresses.

\subsubsection{Spherically symmetric configurations:
Entropy issues
}

\noindent (i) Entropy of spherically symmetric thin shells in vacuum

\noindent The simplest example of a spherically symmetric configuration
is given by a thin shell of radius $R$ surrounded by vacuum, with a
Schwarzschild metric outside and a Minkowski one inside. Its thermal
properties were discussed by Martinez in \cite{mar} but all the examples
studied there do not consider the formation of a quasihorizon. Although
this model of \cite{mar} looks simple, it is of interest in our context,
as it enables to compare the exact results that can be extracted from
the thin shell model with our results, providing thus a convenient test
of our method. Following \cite{mar}, let us write down the first law as
\begin{equation}
TdS=dE+pdA\text{.}  \label{1st}
\end{equation}
Here $E=R\left(1-\sqrt{V}\right)$ is the quasilocal energy \cite{by},
which in this case coincides also with the proper mass of the shell,
$V=1-\frac{r_{+}}{R}$, and $r_{+}=2m$ is the horizon radius, with $m$
being the ADM mass. $A=4\pi R^{2}$ is the surface area of the shell
of radius $R$, and $p$ is the gravitational pressure given by,
\begin{equation}
p=\frac{\left( 1-\sqrt{V}\right) ^{2}}{16\pi R\sqrt{V}}\,,
\end{equation}
see \cite{mar} for details. One can now take two routes. The one
followed by Martinez \cite{mar} and push the thin shell up to the
horizon, and the one advocated by us here. We will see that both routes
give the same result, as they should.

To start, we follow \cite{mar}.
If one takes into account the integrability
conditions of Eq.~(\ref{1st}), and changes variables from $(E,R)$ to
$(m,R)$, it turns out that \cite{mar}
\begin{equation}
T=\frac{T_{0}(m)}{\sqrt{V}},
\label{sm0}
\end{equation}
and Eq.~(\ref{1st}) is reduced to \cite{mar}
\begin{equation}
dS=\frac{dm}{T_{0}(m)}\, . \label{sm}
\end{equation}
Hence the entropy can be found by direct integration. Here, $T_{0}$ has
the usual meaning of the temperature measured by an observer at
infinity. It is seen from (\ref{sm}) that the entropy in this example
does not depend on $R$. This is a consequence of the fact that matter
is absent inside, so $\frac{ \partial S}{\partial R}=0$ everywhere.
Formally, Eq.~(\ref{sm}) is valid everywhere including the near-horizon
region with an arbitrary temperature $T_{0}(m)$. However, near the
horizon
another factor becomes important, which was not taken into account in
\cite{mar} since in it this region was avoided altogether. Outside the
shell there is a backreaction of quantum fields,
and the stress-energy tensor $T_{\mu }^{\nu }$ can be written as
\begin{equation}
T_{\mu }^{\nu }=\frac{T_{0}^{4}-T_{\rm H}^{4}}{g_{00}^{2}}f_{\mu }^{\nu
}+h_{\mu}^{\nu}\,,  \label{set}
\end{equation}
where $f_{\mu }^{\nu }$ and $h_{\mu }^{\nu }$ are finite quantities
(see, e.g., \cite{and} and references therein). At the horizon $g_{00}$
diverges. So, if $T_{0}\neq T_{\rm H}$, inevitable divergences destroy
the horizon of a black hole or the quasihorizon of a quasiblack hole.
Therefore, we must assume that $T_{0}$ is equal to the Hawking
temperature, $T_{0}=T_{\rm H}=\frac{1}{8\pi m} $. Then, the assumption
of negligible backreaction becomes evident and in the main zero-loop
approximation we still may continue to use Eq.~(\ref{sm}). Substituting
$T_{0}=\frac{1}{8\pi m}$ in Eq.~(\ref{sm}), and integrating it, we
obtain $S=4\pi\,m^ 2$, i.e., $S=\pi r_{+}^{2}$, yielding
\begin{equation}
S=\frac14\, A\,,  \label{r2}
\end{equation}
where $A$ now is the quasihorizon area.

Now we follow our formalism and proceed directly from Eq.~(\ref{1st}).
This is less convenient in this simple model, but enables us to check the
general approach. Let us consider variations of the system parameters
for which the shell remains in equilibrium near the would-be horizon,
$R=r_{+}(1+\delta )$ with $\delta$ small and fixed, $0<\delta \ll 1$.
This means that we have to change simultaneously the radius of the shell
$R$ and its ADM mass $m=\frac{r_{+}}{2}$ when we pass from one
equilibrium configuration to another. As a result, the quantity
$V=1-\frac{r_{+}}{R}$ appearing in the expression (\ref{sm}) for the
energy $E=R\left(1-\sqrt V\right)$ is fixed and small. So the first term
in (\ref{1st}) is $dE\approx dR\approx dr_{+}$. The second term in
(\ref{1st}) is huge since $p\sim \frac{1}{\sqrt{V}}\rightarrow \infty$.
Thus, the first term in (\ref{1st}) is negligible as compared to the
second one. Then, writing $p\approx \frac{1 }{16\pi R\sqrt{V}}$,
$T=\frac{T_{0}}{\sqrt{V}}$, $R$ $\approx r_{+}$ and integrating over
$r_{+}$, we reobtain the result (\ref{r2}), where again we
have omitted the integration constant to make sure that $S=0$ when
$r_{+}=0$. This example clearly demonstrates the validity of our
approach and of the key role played by the surfaces stresses, which are
described by the quantity $p$ in this example.

An important and interesting feature can be taken from this model of
Martinez \cite{mar} due to the simplicity of the configuration. The fact
that the entropy $S$ depends on the ADM mass $m$ only, means that the
entropy does not change when the shell with a given $m$ is displaced
radially. In particular, one cannot say that the entropy (\ref{r2}) was
generated in the process of a quasistatic collapse since the entropy
itself remains constant. Its value is due to the value of the Hawking
temperature $T_{\rm H}$. This value for $S$ is enormous when $T_{\rm H}$
is small, which is the case for a configuration with a relatively large
gravitational radius $r_+$. 

Another remark of importance is that it follows from the derivation
given above that no gravitational entropy was assigned to the system a
priori, in accord with some previous observations on the study of thin
shells \cite{page}. Indeed, the Bekenstein-Hawking value for the
entropy (\ref{r2}) was obtained without invoking special additional
assumptions, it is a direct consequence of the first law for matter
only. One can say that ordinary matter in a nontrivial way mimics the
thermal properties of the horizon when its size approaches the
gravitational radius.  In doing so, the requirement $T=T_{\rm H}$ is
essential. One can take $T_{0}\neq T_{\rm H}$ as it was done in
\cite{mar} but only when one considers shells far from the would-be
horizon. To relate both situations, we can consider $T_{0}=T_{\rm
H}[1+\varepsilon \chi (m)]$ where $\varepsilon \ll 1$.  Then, the
backreaction described by Eq.~(\ref{set}) is still bounded, and even
small due to the smallness of $f_{\mu }^{\nu }$ and $h_{\mu }^{\nu }$,
if the numerator in the first term is of the same order of the
denominator, which includes the square of the Schwarzschild metric
coefficient, i.e., $g_{00}^2=(1-\frac{r_{+}}{r})^2$. Then, we obtain
that the admissible minimum radius for the shell is $R=r_{+}(1+\delta
)$ where $\delta \sim \sqrt{\varepsilon}$. The sign of the correction
term to the entropy as compared to (\ref{r2}) depends then on that of
$\chi$.

\vskip 0.4cm
\noindent (ii) Entropy of spherically symmetric continuous
distributions of matter

\noindent
To make the example more realistic, we consider a continuous
distribution of matter, rather than a thin shell. Then, the arguments of
\cite{mar} do not apply and both the entropy and temperature may depend
not only on the mass $m $ but also on radius $R$ of the boundary.

Consider then a general metric for a spherically symmetric distribution of
matter. The metric potentials in Eq.~(\ref{m}) can be chosen such that $
N^{2}(r)=V(r)\,\mathrm{e}^{2\psi (r)}$, $dl=dr/\sqrt{V(r)}$, with $
V(r)\equiv 1-{2m(r)}/{r}$, and $g_{\theta \theta }=g_{\phi \phi }/\sin
^{2}\theta =r^{2}$, with 
$m(r)$ and $\psi (r)$ being the new functions. The
metric is then written in the convenient form
\begin{equation}
ds^{2}=-\left( 1-\frac{2m(r)}{r}\right) \mathrm{e}^{2\psi (r)}dt^{2}+\frac{
dr^{2}}{1-\frac{2m(r)}{r}}+r^{2}(d\theta ^{2}+\sin ^{2}\theta d\phi ^{2})\,,
\label{mv}
\end{equation}
where $m(r)$ and $\psi (r)$ are the relevant metric functions which depend
on the coordinate $r$ alone. Defining $\rho $ and $p_{r}$ as the matter
energy density and radial pressure, respectively, Einstein's equations yield
\begin{equation}
m(r)=4\pi \int_{0}^{r}d\bar{r}\,\bar{r}^{2}\rho \text{,}
\end{equation}
\begin{equation}
\psi (r)=4\pi \int_{r_{0}}^{r}d\bar{r}\,\frac{(\rho +p_{r})\,\bar{r}}{1-
\frac{ 2m(\bar{r})}{\bar{r}}}\,,  \label{uv}
\end{equation}
and a third equation for the tangential pressures that we do not need in our
discussion. Here the constant $r_{0}$ defines the low limit of integration.
In general, it is arbitrary, but there are convenient choices. If the matter
is constrained to the region $r\leq R$, with $\rho+p_{r}=0$ for $r\geq R$,
then one can choose $r_{0}=R$ and $\psi=0$ for $r\geq R$. For example, for
an electrically charged object, one has the Reissner-Nordstr\"{o}m metric
outside in which case indeed $\rho+p_{r}=0$ \cite{vis0}, for all $r\geq R$,
so $\psi=0$ and $m(r)=m-\frac{Q^{2}}{2r}$ where $m$ is the ADM mass and $Q$
is the electric charge. For an electrical neutral object $\rho+p_{r}=0$
trivially and the choice $\psi=0$, $m(r)=m$ for the outside also follows,
yielding the Schwarzschild metric. Of course, other long-range fields, like
dilatonic and other fields, can also be present. If one requires that the
metric is asymptotically flat at infinity, then a good choice is always $
m(r)\rightarrow m$ and $\psi \rightarrow 0$ there, with $r_{0}=\infty $ in
(\ref{uv}). We will see that the final physical results do not depend on the
choice of $r_{0}$. Moreover, we recall that, actually, this approach
works even without requirement of spherical symmetry or asymptotic flatness
as it follows from Sec.~IIA1 where only local properties of the
quasihorizon were used.

Now, for spherically symmetric systems, the first law in terms of the ADM
mass $m$ and boundary radius $R$ has the form, alternative to the form given
in Eq.~(\ref{fl}),
\begin{equation}
T_{0}\,dS=\exp {\psi (R)}\,\left( dm+4\pi p_{r}R^{2}dR\right) \,,
\label{new1stlaw}
\end{equation}
with $p_{r}$ being the radial pressure, see the end of the subsection for 
the
proof of this nontrivial result. Without knowing the system details, one
cannot find in general $S(m,R)$. However, at the quasihorizon limit one can
bypass this restriction. Indeed from Eq.~(\ref{new1stlaw}) we are able to
deduce the Bekenstein-Hawking law for spherical quasiblack holes. The steps
are as follows: (1) Since we want $R\rightarrow r_{+}$ we also have
to put $T_{0}$ as $T_{0}\rightarrow T_{\mathrm{H}}$. Now, $T_{\mathrm{H}}$
is given by $T_{\mathrm{H}}=\frac{\mathrm{e}^{\psi (r_{+})}}{4\pi }\frac{
d(1- \frac{2m(r)}{r})}{dr}(r_{+})$, i.e., at the quasihorizon,
\begin{equation}
T_{0}=T_{\mathrm{H}}=\frac{\mathrm{e}^{\psi (r_{+})}}{4\pi\,r_{+}}\left(
1-8\pi \rho (r_{+})r_{+}^{2}\right) \,.  \label{hawk2}
\end{equation}
Thus, substituting Eq.~(\ref{hawk2}) in Eq.~(\ref{new1stlaw}) we obtain
\begin{equation}
dS=\exp (\psi (R)-\psi ({r_{+}}))\,\frac{dm+4\pi p_{r}RdR}{(1-8\pi \rho
_{+}r_{+}^{2})}\,,  \label{ds2}
\end{equation}
giving the change of the entropy in terms of the changes of the ADM mass and
the boundary radius $R$. (2) Since we are in the quasihorizon limit $
R\rightarrow r_{+}$, the factor $\exp (\psi (R)-\psi ({r_{+}}))$ in
Eq.~(\ref{ds2}) drops out and one has
\begin{equation}
dS=\frac{dm+4\pi p_{r}RdR}{(1-8\pi \rho _{+}r_{+}^{2})}\,,  \label{ds3}
\end{equation}
a simplified version of Eq.~(\ref{ds2}). 
Note that the entropy is a function of $R$ since now
$\frac{\partial S}{\partial R}\neq0$, unlike the case studied
previously, see Eq.~(\ref{sm}).
(3)-(a) Neglecting the difference
between a quasihorizon and a horizon, it follows
\begin{equation}
p_{r}(r_{+})=-\rho (r_{+})\text{,}  \label{prr}
\end{equation}
from the regularity conditions on the horizon itself \cite{vis} (see
also \cite{fn}). (3)-(b) In general, the variations $dm$ and $dR$ are
independent.  However, as we are interested in the quasihorizon limit,
we want to move along the line
\begin{equation}
R\approx 2m=r_{+}\text{, }  \label{Rm}
\end{equation}
in the space of parameters, so that
\begin{equation}
dm\approx \frac{dr_{+}}{2}\,,\;\;dR\approx dr_{+}.  \label{dRdm}
\end{equation}
Thus, putting (\ref{prr})-(\ref{dRdm}) into (\ref{ds3}), yields
\begin{equation}
dS=2\pi r_{+}dr_{+}\,,
\end{equation}
immediately. (4) Then upon integration one recovers, up to a constant $c$
(which one can put to zero), the Bekenstein-Hawking value,
\begin{equation}
S=\frac{1}{4}\,A\,,
\end{equation}
as promised.

Now, we discuss how the two forms of the first law Eqs.~(\ref{fl}) and
(\ref{new1stlaw}) are equivalent. Equation (\ref{fl}) involves the
tangential pressures, whereas Eq.~(\ref{new1stlaw}) the radial
pressure. One relates to quasilocal energy $E$ and the other to the
ADM mass $m$, one to the local temperature $T$ and the other to the
temperature at infinity $T_0$. Consider a thermal compact body at
temperature $T_{0}$ at infinity, with matter with density $\rho$
distributed up to radius $R$, and vacuum outside. For simplicity, we
assume that the shell is massless, so the mass is continuous on the
boundary, $m(R)=m$. Then, Eq. (\ref{fl}) reduces after integration
over angles to
\begin{equation}
TdS=dE+8\pi \Theta _{\theta }^{\theta }RdR  \label{1can1}
\end{equation}
From Eqs.~(\ref{sab})-(\ref{p}), with $\Theta _{ab}^{0}$ 
corresponding to a flat metric, it follows
\begin{eqnarray}
8\pi \Theta _{\theta }^{\theta }= \frac{1}{R}
\left(\sqrt{1-\frac{2m}{R}}
-1\right)+ \left(\frac{1}{2}\frac{1}{\sqrt{1-\frac{2m}{R}}} \frac{
d\left(1-\frac{2m}{R}\right)}{dr}+\sqrt{{1-\frac{2m}{R}}}\;
\frac{d\psi }{dr}
\right)_{r=R_-}\,,  \label{thetaagain}
\end{eqnarray}
and the same for $8\pi \Theta _{\phi }^{\phi }$, where $r=R_-$ means the
quantities are evaluated at $R$ from the inside, and $\psi(R)$ is as in
(\ref{uv}). Now we use the quasilocal energy formula for the spherically
symmetric case \cite{y1,y2,oz},
\begin{equation}
E(R)=R\left(1-\sqrt{1-\frac{2m}{R}}\right)\,.  \label{quasile}
\end{equation}
Performing $dE$ in (\ref{quasile}) and putting it together with 
(\ref{thetaagain}) and $T=T_0/(\sqrt{V}\mathrm{e}^\psi)$ 
(see Eq.~(\ref{tol})) in
Eq.~(\ref{1can1}) yields
\begin{equation}
T_{0}\,dS=\exp{\psi(R)}\,\left(dm+4\pi p_{r}R^{2}dR\right)\,,
\label{new1stlawagain}
\end{equation}
which is precisely Eq.~(\ref{new1stlaw}). We have now completed the
derivation of the equivalence between the two different formulas of the
first law of thermodynamics, Eqs.~(\ref{fl}) and (\ref{new1stlaw}), in a
thorough manner, leaving no doubts about its validity. Note, that 
Eq.~(\ref{new1stlaw}) 
is valid if, in addition to matter, there is also a true black
hole horizon at some radius $r_{\mathrm{bh}}$, in this case the formula for
the mass being slightly changed,
\begin{equation}
m(r)=\frac{r_{\mathrm{bh}}}{2}+ 4\pi\int_{r_{+}}^{r}d\bar{r}\,\bar{r}
^{2}\rho \text{.}
\end{equation}

\vskip0.4cm 
\noindent (iii) Continuity of the entropy function as the radius
approaches the quasihorizon

\noindent In the above consideration, we were interested in obtaining
the asymptotic form of the entropy when the boundary of a body
approaches the quasihorizon, so that we considered the change of the
system configuration along the curve that is approaching the line
$R=2{m}$. On the other hand, it is also important to trace what
happens in the physically relevant situation when the boundary of the
body with fixed ADM mass $m$ changes slowly its position from infinity
toward the horizon, while the radius of the body and its proper mass
are being changed. In the space of parameters $(R,m)$, this
corresponds now to a vertical line $m=\mathrm{const}$. One may ask
what happens to the entropy function in this process near the
quasihorizon, whether a jump in $S(R)$ can occur or not. It follows
from the first law (\ref{ds2}) that $\left( \frac{\partial S}{\partial
R}\right) _{m}=4\pi R^{2}p_{r}(R)$ is finite. Thus, on the
quasihorizon in the process of slowly compressing the shell toward
its own gravitational radius there is no jump in the entropy. This no
jump can be generalized to metrics not necessarily spherically
symmetric.

\subsection{Entropy in the extremal case}

Here we discuss the issue of the entropy for extremal quasiblack
holes. It was argued that $S=0$ in \cite{haw}, see also
\cite{teit,gk}. In \cite{and} it was shown that one has to take into
account that one-loop consideration may change the picture
drastically. So, the issue remains contradictory even in general
relativity. It was also demonstrated within string theory that $
S=A/4$ (see \cite{fn} for a concise review). Because of these
contradictory results, we find it useful to examine the issue of the
entropy of extremal quasiblack holes at the classical level, hoping to
give more insight into it. However, we do not intend to find a
definitive conclusion about the true value of $S$ in this extremal
case. Rather, we only examine which consequences follow from the
assumptions of \cite{haw} when one uses the quasiblack hole picture.

By the definition of the extremal case, $N\sim \exp (Bl)$ where $B$ is
a constant and $l\rightarrow -\infty $. As a result, $\frac{\partial
N}{ \partial l}\sim N$ and we have an additional factor $N\rightarrow
0$ in the numerator in Eq.~(\ref{p}). Therefore $d(s\sqrt{g})=0$ and,
again omitting a constant, we obtain $S=0$. Thus, using the picture of
a thermal body with the boundary approaching its own quasihorizon we
obtain the value $S=0$, with an arbitrary temperature $ T_{0}$, thus
confirming the conclusions of \cite{haw}, see also \cite {gk,teit}.
However, considering that $T_{0}$ is not arbitrary might lead to
another result. The discussion above for the choice of the constant
also holds for extremal quasiblack holes, so $c=0$.

\section{Conclusions}

We have considered the entropy for a system in which a black hole event
horizon never forms, instead a quasihorizon appears. From this we can
draw some remarks:

\noindent (i)  The crucial difference between the usual way of obtaining
the entropy of black holes by integration of the first law and our
version of obtaining the entropy of quasiblack holes, also by
integration of the first law, consists of the fact that we are dealing
with systems which do not have a horizon. Quasiblack holes do not have
horizons as black holes do. The would-be horizon appears only
asymptotically. Thus, as it is exhaustively shown in our paper, it was
not obvious in advance how to get the universal term A/4, which is
intimately connected with a horizon, from matter configurations with
time-like boundaries, instead of a light-like surface as is the case for
a black hole. Our work provides the bridge between thermal matter
configurations and black holes in what concerns entropy. In our view,
this is a very important point.

\noindent (ii) The entropy comes from the quasihorizon surface alone,
i.e., the entropy of a quasiblack hole stems from the contribution of
the states living in a thin layer. That the entropy comes from the
quasihorizon surface alone automatically emphasizes that the properties
of matter inside the quasihorizon are irrelevant, and the final answer for
the entropy is insensitive to them. So, a quasiblack hole deletes
information revealing its similarity to what happens in black hole
physics. Thus, the present work, along with our previous papers on the
mass formula for quasiblack holes \cite {static,statio}, confirms that,
for outer observers, quasiblack holes are objects that yield a smooth
transition to black holes. In particular there is the special
interesting issue of a detailed comparison of the quasiblack hole
picture with the membrane paradigm \cite{membr,pw}.

\noindent (iii) Another important point consists of the role of the huge
surface stresses appearing due to the presence of a quasiblack hole. We
showed that the fact that these stresses are infinite leads to the
Bekenstein-Hawking value $S=A/4$ for the entropy of a nonextremal
quasiblack hole. In doing so, we obtained this result in a
model-independent way and showed that all corrections to the $A/4$ term
vanish in the limit under discussion. For extremal quasiblack holes,
assuming a finite arbitrary temperature at infinity leads to $S=0$ at
the classical level, the fact that the stresses are finite playing a key
role in this derivation. However, considering that  $T_{0}$ is not
arbitrary might lead to another result.

\noindent (iv) The fact that our approach reveals the key role played
by the surface layer near the quasihorizon (which is inevitable there in
the nonextremal case and may appear in the extremal one) supports the
viewpoint according to which the quantum states which generate entropy
live on the quasihorizon of a quasiblack hole. We did not consider here
quantum properties of the system explicitly. However, there is one
important implicit exception. We have assumed that the temperature of
the environment tends to the Hawking value $T_{\mathrm{H}}$. This is a
separate problem that requires further discussion. In particular,
quantum backreaction drastically changes the whole picture in the
extremal case since a non-zero temperature due to backreaction effects
makes the stress-energy tensor of the quantum fields diverge on the
horizon.

\noindent (v) Attempts to place the degrees of freedom that yield the
entropy of a pure black hole on the vicinity of the horizon are not new. One
of those first tries, where the degrees of freedom are on the matter, was
developed in \cite{zt}, while attempts to place the degrees of freedom on
the horizon properties, yielding an entropy coming from the gravitational
field alone, have also been performed, e.g, in \cite{carl} (see \cite{lemos}
for a review). Our approach for the entropy of quasiblack holes shows that
their entropy, although in the matter before its boundary achieves the
quasihorizon, comes ultimately from both the spacetime geometry and the
fields in the local neighborhood at the Hawking temperature. Thus our
approach gives a tie between spacetime and matter in what concerns the
origin of a quasiblack hole entropy. Pushing the analogy between quasiblack
holes and black holes to the end, our approach hints that the black hole's
degrees of freedom appear as non-trivial interplay between gravitational and
matter fields.

\noindent (vi) It would certainly be of further interest to trace the
dynamical process of entropy formation in quasiblack hole scenarios \cite
{bar}.

\noindent (vii) Another important task is the generalization of the
present results to the rotating case.

\begin{acknowledgments}
This work was partially funded by Funda\c c\~ao para a Ci\^encia e
Tecnologia (FCT) - Portugal, through Projects PPCDT/FIS/57552/2004
and PTDC/FIS/098962/2008.
\end{acknowledgments}

\end{document}